 \documentclass[final,5p,times,twocolumn]{elsarticle}

 \biboptions{sort&compress}

\makeatletter
\def\ps@pprintTitle{%
 \let\@oddhead\@empty
 \let\@evenhead\@empty
 \def\@oddfoot{}%
 \let\@evenfoot\@oddfoot}
\makeatother

\usepackage{graphicx}
\usepackage{capt-of}
\usepackage{amsmath}

\usepackage{amssymb}
\usepackage{relsize}
\usepackage[colorlinks,breaklinks]{hyperref}
\usepackage{xcolor}

\usepackage{multirow}

\usepackage{comment}

\newcommand{\bra}{\langle}
\newcommand{\ket}{\rangle}
\newcommand{\bs}[1]{\ensuremath{\boldsymbol{#1}}}

\newcommand{\be}{\begin{equation}}
\newcommand{\ee}{\end{equation}}
\newcommand{\bea}{\begin{eqnarray}}
\newcommand{\eea}{\end{eqnarray}}

\DeclareMathOperator*{\SumInt}{%
\mathchoice%
  {\ooalign{$\displaystyle\sum$\cr\hidewidth$\displaystyle\int$\hidewidth\cr}}
  {\ooalign{\raisebox{.14\height}{\scalebox{.7}{$\textstyle\sum$}}\cr\hidewidth$\textstyle\int$\hidewidth\cr}}   
  {\ooalign{\raisebox{.2\height}{\scalebox{.6}{$\scriptstyle\sum$}}\cr$\scriptstyle\int$\cr}}
  {\ooalign{\raisebox{.2\height}{\scalebox{.6}{$\scriptstyle\sum$}}\cr$\scriptstyle\int$\cr}}  
}

\begin{document}

\begin{frontmatter}

\title{Improved estimates of the nuclear structure corrections in $\mu$D }

\author[Triumf,Manitoba]{O. J.~Hernandez}
\ead{javierh@triumf.ca}

\author[Triumf]{C.~Ji}
\ead{jichen@triumf.ca}

\author[Triumf,Manitoba]{S.~Bacca}
\ead{bacca@triumf.ca}

\author[Hebrew]{N.~Nevo Dinur}
\ead{nir.nevo@mail.huji.ac.il}

\author[Hebrew]{N.~Barnea}
\ead{nir@phys.huji.ac.il}

\address[Triumf]{TRIUMF, 4004 Wesbrook Mall, Vancouver, BC V6T 2A3, Canada}
\address[Manitoba]{Department of Physics and Astronomy, University of Manitoba, Winnipeg, MB, Canada R3T 2N2}
\address[Hebrew]{Racah Institute of Physics, The Hebrew University, Jerusalem 9190401, Israel}


\begin{abstract}
We calculate the nuclear structure  corrections to the Lamb shift in muonic deuterium
by using state-of-the-art nucleon-nucleon potentials derived from chiral effective field theory.
Our calculations complement  previous theoretical work
obtained from phenomenological potentials and the zero range approximation. The study of the chiral
convergence order-by-order and 
the dependence on cutoff variations
allows us to improve the estimates on 
the nuclear structure corrections and the theoretical uncertainty
coming from nuclear potentials. This will enter the determination of the nuclear radius from 
ongoing muonic deuterium experiments at PSI.
\end{abstract}


\begin{keyword}
muonic atoms \sep nuclear polarizability \sep chiral potential
\end{keyword}

\end{frontmatter}

\section{Introduction}

The root-mean-square charge radius of the proton
 was recently determined by spectroscopic measurements of the 2S-2P atomic shift, 
 {\it i.e.}, the Lamb shift (LS)~\cite{Lamb47},
 in muonic hydrogen~\cite{Pohl10,Antognini13}, where the proton 
is orbited by
a muon instead of 
an electron
 as in ordinary hydrogen. 
With respect to the CODATA-2010 compilation~\cite{CODATA}, which is
based on the combined electron proton scattering data and the spectroscopic measurements in the ordinary hydrogen atom, 
the accuracy was improved ten-fold and a proton radius value smaller by 7 standard deviations was observed. 
This large deviation between the muonic and the electronic measurements
constitutes the so-called ``proton radius puzzle''. 
It has attracted a lot of attention since 2010, from both theoretical and experimental viewpoints. 
Several beyond-the-standard-model theories, including lepton universality violations, have attempted to solve this puzzle (see {\it e.g.}~\cite{Pohl13} for a review).
For example, the authors of Refs.~\cite{Batell11, TuckerSmith11, CarlsonRislow14} investigated the possibility of the existence of new interaction mediators that can explain not only the proton radius puzzle,
but also the  $(g-2)$ muon anomaly.
As yet, none of these theories 
have been either verified or ruled out by experiments.
Alternative explanations are being 
sought after either through novel aspects of hadronic structure~\cite{Miller11, 
Birse12, Jent13}, 
or from renewed analyses of the electron scattering 
 data, {\it e.g.}, Refs.~\cite{Lorenz12,Kraus14}. 
To date, no commonly accepted explanation of the puzzle has been found. 
Various new dedicated experiments have been planned to measure electron~\cite{JLAB,Mainz}
and muon~\cite{MUSE} scattering on the proton. 
In addition, experimental reexamination of ordinary hydrogen spectroscopy is under way, {\it e.g.}, Ref.~\cite{Beyer13}.
A complementary experimental program based on high-precision spectroscopic 
measurements on various muonic atoms aims to study the systematics of the
discrepancy with ordinary atoms as a function of the
atomic mass A and charge number Z.
In particular, the CREMA collaboration~\cite{Antognini11} plans to measure the 
 Lamb shift
and isotope shifts in several light muonic atoms. 
The deuteron is the 
 lightest
compound nucleus, 
 made up of one proton and one neutron, 
 and it plays an important role in few-body nuclear physics.
The Lamb shift of its muonic atom, $\mu$D, 
is currently being measured at PSI. 
 This measurement 
will provide a solid and independent test of the systematic uncertainties in the $\mu$H experiment. 
Furthermore, assuming a new interaction mediator 
 that violates 
lepton 
 universality,
the $\mu$D experiment 
 may help to constrain the possible couplings of this new interaction to the proton and the neutron. 
Therefore, it is important to compare the deuteron charge radius
$\bra r_{ch}^2\ket^{1/2}_d$ 
extracted from the $\mu$D Lamb shift
with the values determined from previous and ongoing experiments on
$e$D scattering~\cite{Sick98,eD_Mainz}, as well as from the precision measurements on
the H/D isotope shift~\cite{Parthey10,Jentschura11}.

The extraction of the nuclear charge radius 
from LS
measurements relies heavily on theoretical input. 
For the deuteron, 
the 2S-2P energy transition is related to $\bra r_{ch}^2\ket_d^{1/2}$ 
by~\cite{Friar79} 
\begin{equation} 
\label{eq:E2s2p}
\Delta E  =
 \delta_{QED}+\delta_{\rm pol}+\delta_{\rm Zem}+
 \frac{m_r^3 \alpha^4}{12} \bra r_{ch}^2\ket_d~,
\end{equation}
where
$\alpha$ is the fine-structure constant and $m_r$ is the reduced muon mass. 
Quantum electrodynamic (QED) corrections $\delta_{QED}$, 
as well as nuclear structure corrections $\delta_{\rm pol}$ and $\delta_{\rm Zem}$, 
are obtained from theoretical calculations. 
 $\delta_{QED}$, which originates from 
vacuum polarization, lepton self-energy, and relativistic recoil effects, 
is known with very good accuracy~\cite{Borie12}. 
The uncertainty of the extracted radius 
is by far limited by the uncertainty of the nuclear structure corrections.
These corrections arise from the two-photon exchange (TPE)
(see {\it e.g.}~Fig.1 in Ref.~\cite{Ji13}), 
in which the virtual photons transfer energy and momentum to the nucleus. 
 The TPE contribution is traditionally separated into the sum of the elastic Zemach term
 $\delta_{\rm Zem}$, proportional to the third Zemach moment~\cite{Friar79},
 and the inelastic polarization term $\delta_{\rm pol}$, 
 {\it i.e.}, $\delta_{\rm TPE} = \delta_{\rm Zem}+\delta_{\rm pol}$. 
The inelastic part is further separated into
$\delta_{\rm pol} = \delta^{A}_{\rm pol}+\delta^{N}_{\rm pol}$,
the sum of the nuclear polarization $\delta^{A}_{\rm pol}$ and
the intrinsic nucleon polarizability $\delta^{N}_{\rm pol}$. 

For $\mu$D, these nuclear structure
corrections have been most recently estimated by Carlson {\it et al.}~\cite{Carlson14} using forward dispersion relations to analyze experimental elastic
deuteron form factors and inelastic electromagnetic deuteron scattering data.
Due to lack of data in the most relevant low-energy 
quasielastic
regime, this analysis suffers from a $35\%$ uncertainty.
Theoretical estimates of nuclear corrections
were recently made by Friar~\cite{Friar13}, utilizing the zero-range
model. 
The uncertainty was roughly estimated to be 1-2$\%$ due to missing higher-order corrections, mainly from two-body operators and S-D mixing in the deuteron ground state~\cite{Phillips02}. 
However, this uncertainty is obtained by dimensional analysis,  and is thus only an approximation. 
A pioneering calculation of $\delta_{\rm TPE}$ in $\mu$D was made by Pachucki~\cite{Pachucki11}, who used a high precision nucleon-nucleon (NN)
force of phenomenological nature, namely the AV18 potential~\cite{AV18}. 
Pachucki quoted $\sim 1\%$ uncertainty on $\delta_{\rm TPE}$, mostly due to higher-order atomic-physics terms in the $\alpha$ expansion, but did not include the uncertainty from nuclear potentials. 

In order to obtain an improved estimate,
one may rely on microscopic calculations based on
nuclear Hamiltonians constructed with state-of-the-art NN
potentials~\cite{AV18, Machleidt11, Epelbaum09}.
The NN force is an effective potential emerging from the
low-energy quantum chromodynamics (QCD). As such it depends on 
the adopted resolution scale, a set of fitting parameters, and
truncation orders.
Realistic NN potentials are typically fitted to
reproduce NN scattering data (mostly $np$ and $pp$) with high accuracy
$\chi^2\approx 1$
up to pion production threshold. 
In order to provide a rigorous
estimate of the theoretical 
uncertainty one should consider the effect of varying all the fitting parameters
within a certain confidence interval. Such a task is at the moment out of hand.
Alternatively, one could explore various realistic NN potentials as a sample of
the possible model space. 

An early attempt to estimate the theoretical error of the nuclear polarization in $\mu$D
using several of the NN forces available at that time was 
made by Leidemann and Rosenfelder~\cite{Leidemann95}, who found a potential dependence less than 2$\%$. However contributions from Coulomb distortion
were missing in that calculation.

The purpose of this Letter is to provide new calculations and improved estimates of the theoretical uncertainty in the nuclear structure corrections using state-of-the-art nuclear potentials from chiral effective field theory ($\chi$EFT) ~\cite{Machleidt11, Epelbaum09}. 
The $\chi$EFT nuclear potentials result from 
a systematic expansion of the interaction in powers of 
 a soft momentum scale over a hard momentum scale
$Q/\Lambda_{\chi}$ \cite{Machleidt11}. The soft scale $Q$ is 
associated with the typical momenta of the system or dynamics under consideration, and the hard scale $\Lambda_{\chi}$ 
determines at which scale 
the effective theory breaks down. 
The short range part of the $\chi$EFT potential stems from 
contact terms in the Lagrangian, which are controlled by free parameters in the $\chi$EFT Lagrangian, commonly known as low energy constants (LECs). 
Their values are set 
by fitting the interaction
to the NN scattering data at minimum $\chi^2$/datum for given resolution scales, determined by the regularization cutoff $\Lambda\sim \mathcal{O}(\Lambda_{\chi})$. 
For the chiral forces 
better $\chi^2$/datum 
and smaller $\Lambda$ dependence in predictions
is obtained with 
the increase of the expansion order.
One can improve the description of the $np$ scattering 
data, going from a $\chi^2$/datum of 36.2 at next-to-leading order (NLO) calculation
 up  to 1.1 at next-to-next-to-next-to-leading order (N$^3$LO)  in the energy range of 0-290 MeV~\cite{Entem03}.  
By performing a systematic study in $\chi$EFT order-by-order and by varying the 
resolution scale $\Lambda$,
we can better assess the theoretical error for $\delta_{\rm TPE}$ stemming from 
nuclear forces.

\section{Calculation details}

To calculate the deuteron polarization
we essentially follow the derivation provided in our recent paper~\cite{Ji13}. 
Within the multipole expansion formalism
 we take into account the leading electric dipole term $\delta^{(0)}_{D1}$, 
 the Zemach related contribution $\delta^{(1)}_{Z3}$,
and other multipole corrections including a monopole $\delta^{(2)}_{R2}$, a quadrupole $\delta^{(2)}_{Q}$ and an interference between two rank-1 operators $\delta^{(2)}_{D1D3}$. 
The small parameter in the expansion is $\eta=\sqrt{m_r/m_d}$ where
$m_r$ is the reduced muon mass and $m_d$ is the deuteron mass.
According to our power counting, 
the contribution of a term with superscript 
$(1)$ (or $(2)$) to $\delta_{\rm pol}^A$ is suppressed by $\eta$ (or $\eta^2$)
relative to that of the leading $\delta^{(0)}_{D1}$ term. 
We also include the Coulomb distortion corrections $\delta^{(0)}_{C}$, 
relativistic longitudinal and transverse corrections  $\delta^{(0)}_{L}$ and $\delta^{(0)}_{T}$, magnetic dipole corrections $\delta^{(0)}_{M}$, and finite-nucleon-size corrections including $\delta^{(1)}_{Z1}$, $\delta^{(1)}_{R1np}$ and $\delta_{NS}^{(2)}$.
Therefore, we calculate $\delta_{\rm pol}^A$ as
\begin{eqnarray}
\label{delta}
\delta^A_{\rm pol} &=&
\left[\delta^{(0)}_{D1}+\delta^{(0)}_{C}+\delta^{(0)}_{L}+\delta^{(0)}_{T} +\delta^{(0)}_{M} \right]
+ \delta^{(1)}_{Z3}
\nonumber\\
&& + \left[\delta^{(2)}_{R2}+\delta^{(2)}_{Q}+\delta^{(2)}_{D1D3} \right] 
+ \left[\delta^{(1)}_{Z1} + \delta^{(1)}_{R1np} + \delta^{(2)}_{NS} \right].
\end{eqnarray}
Detailed formulas relating (most of) these corrections 
are found in~\cite{Ji13} and are not repeated here. 
Except for $\delta^{(1)}_{Z3}$, $\delta^{(1)}_{Z1}$ and $\delta^{(1)}_{R1np}$, which are ground-state observables, each of the above contributions
can be written as a sum of terms of the form
\begin{equation}\label{eq:SR}
\delta_{a}=\int_{\omega_{\rm th}}^{\infty} d\omega\, S_a\left(\omega\right)g_a\left(\omega\right),
\end{equation}
where $g_a\left(\omega\right)$ is an energy-dependent weight function 
(different for each of the corrections),
$\omega_{\rm th}$ is the threshold excitation energy,
 and $S_a(\omega)$
is the response function.
$S_a(\omega)$
is given by
\begin{equation}
\label{eq:S_omg}   
   S_a(\omega)=
   \SumInt\limits_{f\neq 0} |\bra f |\hat{O}_a| 0 \ket|^2
    \delta\left(E_{f}-E_{0}-\omega\right)
\end{equation}
where
$\hat O_a$ is the transition operator (that can be different in the above corrections), $|0\ket$ and $|f\ket$ are the ground and final
eigenstates of the Hamiltonian $H$ with eigenvalues $E_0$ and $E_f$ respectively. 
The integration  in Eq.~(\ref{eq:S_omg}) carries over the excited scattering states of the deuteron.

The Coulomb corrections $\delta^{(0)}_C$ used here contain only the term of order $\alpha^6\ln\alpha$. Therefore we have
\begin{equation}
\label{eq:delta_C}
\delta^{(0)}_C
= -\frac{2\pi}{9} m_r^3 \alpha^6 \int_{\omega_{th}}^{\infty} d\omega\, \frac{m_r}{\omega} \ln\frac{2 m_r \alpha^2 }{\omega}\, S_{D1}(\omega)~,
\end{equation}
where $S_{D1}(\omega)$ is the electric-dipole response function with operator $D_1\equiv R_p Y_{1}(\hat{R}_p)$~\cite{Ji13}. The terms of higher orders in $\alpha$, which were included in $\delta^{(0)}_C$ of Ref.~\cite{Ji13}, only correct $\delta^A_{\rm pol}$ by $\lesssim 0.25\%$, and are thus neglected in this analysis.

Because the deuteron is a nucleus with the total angular momentum $J_0=1$ in the ground state, the magnetic term $\delta^{(0)}_M$,
which is negligible in the $\mu^4{\rm He}^+$, has to be included here. It relates to the magnetic response function $S_{M1}(\omega)$ by
\begin{equation}
\label{delta_m1}
\delta^{(0)}_M = \frac{1}{3} m_r^3 \alpha^5 
\left(\frac{g_p-g_n}{4 m_p}\right)^2
\int^{\infty}_{\omega_{\rm th}} d\omega \sqrt{\frac{\omega}{2 m_r}} S_{M1} (\omega)~, 
\end{equation}
where $g_p=5.586$, $g_n=-3.826$, and $m_p$ is the proton mass. The magnetic dipole (M1) operator is defined by $\vec{s}_p - \vec{s}_n$~\cite{Pachucki11}.

The sum of the terms $\delta^{(1)}_{Z3}$ and $\delta^{(1)}_{Z1}$ cancels exactly 
the elastic Zemach term $\delta_{\rm Zem}$:
\begin{equation}
\label{eq:Zemach}
\delta_{\rm Zem} = - \left[ \delta^{(1)}_{Z3} + \delta^{(1)}_{Z1} \right]
= -m_r^4 \alpha^5 \left[\frac{1}{24}\, \bra r^3\ket_{(2)} + \left(\frac{2}{\beta^2}-\lambda \right)\, \bra r\ket_{(2)} \right],
\end{equation}
where $\beta=4.120$ fm$^{-1}$ and $\lambda= 0.01935$ fm$^2$ are used as in Refs.~\cite{Friar13,Ji13} to reproduce the proton charge radius $\bra r_{ch}^2 \ket_p^{1/2} = 0.8409$ fm~\cite{Antognini13} and the neutron charge radius squared  
$\bra r_{ch}^2 \ket_n = -0.1161$~fm$^2$ \cite{Beringer:1900zz}.  The $n$-th moment $\bra r^n \ket_{(2)}$ is defined by
\begin{equation}
\bra r^n \ket_{(2)} \equiv \iint d\bs{R} d\bs{R}' \rho_p(\bs{R})\,  |\bs{R}-\bs{R}'|^n\, \rho_p(\bs{R}')~,
\end{equation}
with $\rho_p$ denoting the point-proton density in deuteron. The cancellation of $\delta_{\rm Zem}$ was explicitly applied in Refs.~\cite{Pachucki11, Friar13}, whose results for nuclear structure effects are provided as the combination of $\delta_{\rm pol}^A$ and $\delta_{\rm Zem}$. In this work, we will also provide our final result as a combination of the two.

Unlike the $\mu^4{\rm He}^+$ case of Ref.~\cite{Ji13}, a $pp$ charge correlation is not present in $\mu$D, as it contains only one proton. 
On the other hand, $\delta^{(1)}_{R1np}$, which denotes a contribution from the $np$ charge correlation, has to be included, and is calculated as
\begin{equation}
\label{eq:R1np}
\delta^{(1)}_{R1np} = \lambda m_r^4\alpha^5 \bra 0 |\, |\bs{R}_p -\bs{R}_n|\, | 0\ket~,
\end{equation}
where the operator $|\bs{R}_p -\bs{R}_n|$ denotes the relative distance between the proton and the neutron.

To calculate $|0\ket$ and $|f\ket$ we expand them on
the harmonic oscillator (HO) basis~\cite{messiah}
and then diagonalize
the Hamiltonian matrix, whose dimension is set by $N_{max}=2n+\ell$, where $n$ is the HO quantum  number and $\ell$ is the relative angular momentum between the proton and the neutron. We use the HO basis because it is very flexible and allows the use of both local potentials in coordinate space, like the AV18, as well as non-local forces, like those derived in $\chi$EFT, within the same framework.  From the diagonalization of the Hamiltonian matrix we  
obtain a set of eigenstates $|\mu\ket$ and eigenvalues $E_\mu$. 
Once the size of the basis is large enough, the lowest eigenvalue state coincides with 
the ground-state of the deuteron. All the other states
$|\mu\ket$  can be regarded as a discretization of the two-body continuum.
In terms of this discrete basis, Eq.~(\ref{eq:SR}) becomes
a sum
over the transition probabilities from the
ground state to the discretized excited states with weighted functions $g_a$ as
\begin{equation}
\label{I_sum}
  \delta_a 
   =\sum\limits_{\mu \neq 0}^{N_{max}} |\bra \mu | \hat O_a | 0\ket|^2 g_a(\omega_{\mu})\;,
\end{equation}
where $\omega_{\mu}=E_{\mu}-E_0$ is the excitation energy, and the finite 
sum runs over all the calculated states.
As proven in Ref.~\cite{Nevo14}, even if one is approximating continuum states with discrete states,
Eq.~(\ref{I_sum}) becomes exact when the size of the basis becomes large enough (with increasing $N_{max}$). 
Furthermore, for the two-body problem the basis size remains tractable so that a direct diagonalization is always possible and one 
does not need to introduce the Lanczos algorithm as in Ref.~\cite{Nevo14}.

We calculate the deuteron ground-state properties and polarizability using the AV18 potential~\cite{AV18} and NN forces from $\chi$EFT. 
The AV18 results serve as checks with Ref.~\cite{Pachucki11} 
and the $\chi$EFT results are genuine new. We use
chiral potentials developed by Epelbaum {\it et al.}~\cite{Epelbaum05} at different order in $\chi$EFT, 
starting from NLO to N$^3$LO. For these, a certain range of cutoff variation will be 
explored, going from 
$\Lambda=400$ to 600 MeV for the 
cutoff to regularize the Lippmann-Schwinger equation~\cite{Lippmann-Schwinger}
and from $\tilde{\Lambda}=600$ to 700 MeV for the spectral function cutoff (see \cite{Epelbaum05} for details). Hereafter, we shall refer to these potentials
either individually as N$^k$LO$(\Lambda,\tilde\Lambda)$ or collectively as N$^k$LO-EGM.
 We will also use the chiral potential at N$^3$LO developed by Entem and Machleidt at a fixed cutoff $\Lambda=500$ MeV~\cite{Entem03} (N$^3$LO-EM). By doing so, we will provide a systematic study of the convergence and control the theoretical uncertainty.   

\section{Results}

\begin{figure}[htb]
\centerline{
    {\includegraphics[angle=0,width=0.9\linewidth]{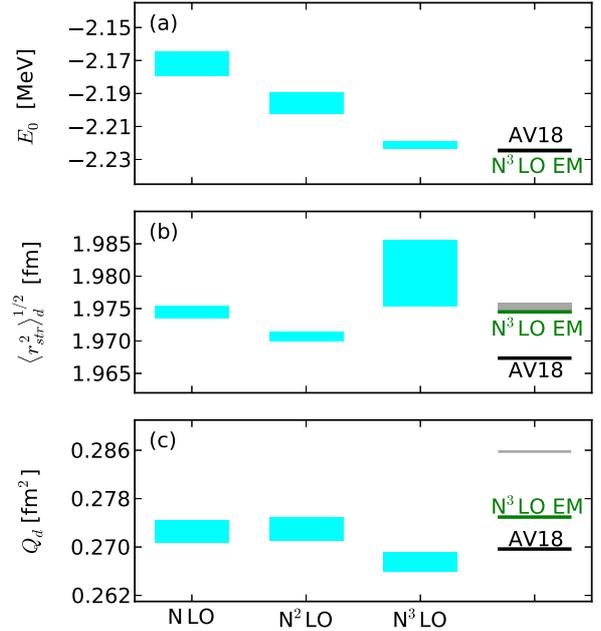}}}
\caption{Deuteron energy (a), LO structure radius (b) and quadrupole moment (c)  at different
chiral orders for the various choices of the cutoffs $\{\Lambda, \tilde{\Lambda}\}$.  Compared are the results with the AV18 and N$^3$LO-EM potentials and the experimental values (gray band).
} 
\label{fig_1_gs} 
\end{figure}

To estimate
the possible spread in $\delta_{\rm TPE}$ 
due to the 
variation in nuclear Hamiltonian models, 
we first present few deuteron ground-state
observables and follow their evolution with the nuclear forces.
Our numerical results for 
the energy $E_0$, the root-mean-square 
structure radius $\bra r_{str}^2 \ket_d^{1/2}$, 
the quadrupole moment $Q_d$,
the electric 
polarizability $\alpha_E$, and the magnetic susceptibility $\beta_M$
are presented in Table~\ref{table_gs}. The first three are also presented graphically in Fig.~\ref{fig_1_gs}.
With the AV18 and  N$^3$LO-EM we reproduce the experimental binding energy, $ 2.224573(2)$ MeV~\cite{Audi19971},
 up to 5 decimal 
digit  as in Refs.~\cite{AV18} and~\cite{Entem03}. 
For the chiral N$^k$LO$(\Lambda,\tilde\Lambda)$ potentials we use the cutoff values 
provided by Epelbaum~\cite{Epelbaum-14}.
We note that the range of the explored cutoffs  does not coincide exactly with the 
cutoff range spanned in Ref.~\cite{Epelbaum05}. Only for N$^2$LO there is an exact one-to-one correspondence of cutoff sets used; there we 
agree with
Epelbaum's values within the third decimal digit for energy, radius and quadrupole moment.

\begin{table}[htb]
\caption{Deuteron electromagnetic observables, as calculated by different potential models.}
\begin{center}
\footnotesize
\label{table_gs}
\begin{tabular}{l c c c c c}
\hline\hline
  & $E_0$  & $\bra r_{str}^2 \ket_d^{1/2}$ & Q$_d$    & $\alpha_E$ & $\beta_M$ \\ 
  & [MeV] &  [fm]  & [fm$^2$] & [fm$^{3}$] & [fm$^{3}$] \\ 
\hline
NLO(400,700)         & 2.1647 & 1.975 & 0.2707 & 0.652 & 0.0706 \\
NLO(550,700)         & 2.1794 & 1.974 & 0.2745 & 0.647 & 0.0696 \\
N$^2$LO(450,700)     & 2.2022 & 1.970 & 0.2711 & 0.640 & 0.0695 \\
N$^2$LO(550,600)     & 2.1890 & 1.971 & 0.2749 & 0.644 & 0.0694 \\
N$^2$LO(600,700)     & 2.1999 & 1.970 & 0.2747 & 0.640 & 0.0690 \\
N$^3$LO(450,700)     & 2.2189 & 1.986 & 0.2659 & 0.637 & 0.0695 \\
N$^3$LO(550,600)     & 2.2196 & 1.979 & 0.2673 & 0.635 & 0.0693 \\
N$^3$LO(600,700)     & 2.2235 & 1.975 & 0.2692 & 0.633 & 0.0689 \\
N$^3$LO-EM           & 2.2246 & 1.974 & 0.2750 & 0.633 & 0.0684 \\
AV18                 & 2.2246 & 1.967 & 0.2697 & 0.633 & 0.0679 \\
\hline \hline
\end{tabular}
\end{center}
\end{table}

The electromagnetic observables in Table~\ref{table_gs} (all but $E_0$) have been calculated assuming that nucleons are point-like and  using only one-body operators with no relativistic corrections (RC) or meson-exchange currents (MEC). 
The nonrelativistic one-body operators correspond to the dominant contributions to the electromagnetic observables in the $\chi$EFT expansion.

The charge radius is related to the 
structure
radius by finite nucleon size corrections as 
\begin{equation}
\label{eq:rad}
\bra r_{ch}^2 \ket_d=\bra r_{str}^2 \ket_d + \bra r_{ch}^2\ket_p+ \bra r_{ch}^2\ket_n \,, 
\end{equation}
where the structure radius is separated into 
the leading component from the nonrelativistic one-body current and the subleading part from RC+MEC.
In the $\chi$EFT language, RC and MEC (also called two-body currents)
enter at higher order in the $Q/\Lambda_{\chi}$ expansion with respect to 
the leading component,
specifically at N$^2$LO and N$^3$LO, respectively, see {\it e.g.}~\cite{review}.
Similar corrections enter in other electromagnetic observables as well.
Relativistic and two-body  corrections should be evaluated consistently within a theory as to satisfy gauge invariance.
They have already been derived in $\chi$EFT, see {\it e.g.}~\cite{Park03,Pastore08,Kolling09}, where pion-exchange and contact two-nucleon currents  appear, but are not applied in this work.
Instead, following Refs.~\cite{Entem03,AV18},  we show in Table~\ref{table_mec_r} that, adding phenomenologically RC and MEC corrections,  the calculated values for $\bra r_{str}^2 \ket^{1/2}_d$ and $Q_d$ are consistent with the most recent experimental results~\cite{Jentschura11,Pavanello10}.\footnote{In Ref.~\cite{Jentschura11}, $\bra r_{str}^2 \ket^{1/2}_d$ 
is extracted from the H-D isotope shift measurement, which does not explicitly depend on the measurement of $\bra r_{ch}^2\ket_p $.}

\begin{table}[htb]
\caption{Deuteron ground state properties and electric polarizability without and with RC+MEC compared to experimental data.}
\footnotesize
\begin{center}
\renewcommand{\tabcolsep}{1.5mm}
\label{table_mec_r}
\begin{tabular}{l  l l l l}
\hline\hline
  & & $\bra r_{str}^2 \ket^{1/2}_d$ [fm] & Q$_d$ [fm$^2$] & $\alpha_E$ [fm$^{3}$] \\ 
\hline
N$^3$LO-EM~~ & This work     & 1.974 & 0.2750 & 0.633\\
       & +RC+MEC & 1.978 & 0.285$^1$ & \\
\hline                   
\multirow{2}{*}{AV18}         & This work    & 1.967 & 0.2697  & 0.633\\
                              & +RC+MEC    &  $\;\;\;-$ & 0.275$^2$ & \\
\hline
\multicolumn{2}{l}{Experiment}   &  1.97507(78)$^3$ & 0.285783(30)$^4$  
          & 0.70(5)$^5$    
\\
        &  &  & & 0.61(4)$^6$ \\
\hline \hline
\end{tabular}
\end{center}
$^1$Ref.~\cite{Entem03}, $^2$Ref.~\cite{AV18}, $^3$Ref.~\cite{Jentschura11}, $^4$Ref.~\cite{Pavanello10}, $^5$Ref.~\cite{Rodning82}, $^6$Ref.~\cite{Friar83}.
\end{table}

Comparing our $\alpha_E$  calculated with the AV18 potential
with the result of \cite{FriarPayne97} we get an agreement of about 0.15\%.
Our N$^3$LO-EGM and N$^3$LO-EM results agree within 1\% 
among the AV18 result and
the pion-less EFT results of \cite{Chen98}.
For the magnetic susceptibility 
using the AV18 potential we obtain $\beta_M=0.0679$ fm$^3$, 
which agrees within 0.1\% with previous calculations using  the same potential and within 1\% with other calculations using different potentials~\cite{FriarPayne97-2}. The latter is
comparable with the sensitivity to the NN potential that we observed in Table~\ref{table_gs}. In fact,
our N$^3$LO-EGM and N$^3$LO-EM results for 
$\beta_M$ scatter at $\sim 1\%$ and are within few percent of the results based on pion-less EFT from Ref.~\cite{Chen98}.

In Fig.~\ref{fig_1_gs}, we first show  the binding energy. One observes that as the chiral order gets higher, convergence is approached and the band due to the cutoff variation becomes order-by-order smaller.  One also notes, that despite the fact that the deuteron binding energy is a prediction of N$^k$LO-EGM potentials~\cite{Epelbaum05}, it 
agrees well
with the values from the AV18 and N$^3$LO-EM potentials, where the deuteron binding energy has been included in the fit~\cite{Entem03, AV18}.
 Overall, they deviate from the experimental value by only 0.4$\%$.

For the structure  radius  the situation is different. While quite a nice convergence in the chiral order is achieved, the band obtained by the cutoff variation becomes larger  at N$^3$LO. Interpreting the cutoff band as an estimate of the theoretical error, it means the error is larger for the highest chiral order, which also results in a larger deviation from the experimental value.
This was already observed in  Ref.~\cite{Epelbaum05} and attributed to the effect of 
two-body currents.
For $\bra r_{str}^2 \ket_d^{1/2}$ 
and $Q_d$ we find that the values obtained with the chiral potential N$^3$LO-EM and AV18 fall marginally out of the cutoff band spanned by the N$^3$LO-EGM potentials.
This could imply that a slightly different parameterization
of the two-body currents is needed for each potential. 
If we introduced the two-body operators, they would come with new low-energy constants (LECs). 
The new LECs
would have to be calibrated in a way to compensate the cutoff dependence, thus reducing the theoretical errors.

A full compilation of the nuclear structure corrections
is displayed in Table~\ref{table_pol}.
 We first discuss the calculation with the AV18 potential, 
comparing our results with Pachucki ~\cite{Pachucki11}.
As already pointed out in~\cite{Ji13}
we find very good agreement with Pachucki  
for the leading order dipole and Coulomb 
correction terms.
Small but non-negligible differences appear in
the relativistic corrections. These differences can be traced to the
leading-order truncation in the $\omega/m_r$ expansion of the weight function $g_a$
made in~\cite{Pachucki11}, whereas we have used the full form given in~\cite{Ji13}.
Our results for $\delta^{(2)}_{R2}$, $\delta^{(2)}_{Q}$ and $\delta^{(2)}_{D1D3}$ are smaller than those obtained by Pachucki in~\cite{Pachucki11} by $\sim 8\%$, although the same formulas are used.
As an independent check, we use the deuteron electric-quadrupole response function calculated by Arenh\"{o}vel~\cite{Arenhoevel,Arenhoevel-private} from the Paris potential and obtain $\delta_Q=0.060$ meV, which is only $1\%$ smaller than our corresponding value.\footnote{The difference is calculated with $\delta_Q$ in more digits than is presented in Table~\ref{table_pol}.}
This $1\%$ discrepancy is also consistent with the difference of deuteron structure radius between the two calculations and compatible with the sensitivity to different potentials.
Our calculated magnetic dipole contribution $\delta^{(0)}_{M}$ is $\sim 50\%$ smaller than that in~\cite{Pachucki11}, 
despite the fact that the same formula, {\it i.e.}, Eq.~(\ref{delta_m1}), is used. 
As noted above, our values for  
the related magnetic susceptibility $\beta_M$ stand in line with previous calculations. 
We have also integrated the deuteron $M1$ response function from Arenh\"{o}vel~\cite{Arenhoevel,Arenhoevel-private} and obtained $\delta^{(0)}_{M}=0.0067$ meV.

\begin{table}[htb]
\caption{Nuclear polarization contributions to the $2S$-$2P$ Lamb shift
  $\Delta E$ [meV] in $\mu$D with different potentials. }
  \footnotesize
\begin{center}
\label{table_pol}
\begin{tabular}{l c c c c c}
\hline\hline
          &        & Ref.~\cite{Pachucki11} & AV18 & N$^3$LO-EM & N$^3$LO-EGM     \\
\hline
$\delta^{(0)}$ & $\delta^{(0)}_{D1}$   & -1.910 & -1.907 & -1.912 & (-1.911,-1.926) \\
               & $\delta^{(0)}_{L}$    &  0.035 &  0.029 &  0.029 & ( 0.029, 0.030) \\
               & $\delta^{(0)}_{T}$    &    $-$ & -0.012 & -0.012 &  -0.013         \\
               & $\delta^{(0)}_{C}$    &  0.261 &  0.262 &  0.262 & ( 0.262, 0.264) \\ 
               & $\delta^{(0)}_{M}$    &  0.016 &  0.008 &  0.007 &   0.007         \\
$\delta^{(1)}$ & $\delta^{(1)}_{Z3}$   &    $-$ &  0.357 &  0.359 & ( 0.359, 0.363) \\             
$\delta^{(2)}$ & $\delta^{(2)}_{R2}$   &  0.045 &  0.042 &  0.041 &   0.041         \\
               & $\delta^{(2)}_{Q}$    &  0.066 &  0.061 &  0.061 &   0.061         \\
               & $\delta^{(2)}_{D1D3}$ & -0.151 & -0.139 & -0.139 & (-0.139,-0.140) \\
$\delta_{NS}$  & $\delta^{(1)}_{Z1}$   &    $-$ &  0.064 &  0.064 & ( 0.064, 0.065) \\
               & $\delta^{(1)}_{np}$   &    $-$ &  0.017 &  0.017 &   0.017         \\
               & $\delta^{(2)}_{NS}$   &    $-$ & -0.015 & -0.015 &  -0.015         \\
\hline
\multicolumn{2}{l}{$\delta^A_{\rm pol}$}   &    $-$ & -1.235 & -1.237 & (-1.236,-1.246) \\ 
\multicolumn{2}{l}{$\delta_{\rm Zem}$}   &    $-$ & -0.421 & -0.423 & (-0.424,-0.428) \\                                     
\multicolumn{2}{l}{$\delta^A_{\rm pol}+\delta_{\rm Zem}$}
                                       & -1.638 & -1.656 & -1.661 & (-1.660,-1.674) \\                     
\hline\hline
\end{tabular}
\end{center}
\end{table}

\begin{figure}[htb]
\centerline{
    {\includegraphics[angle=0,width=0.9\linewidth]{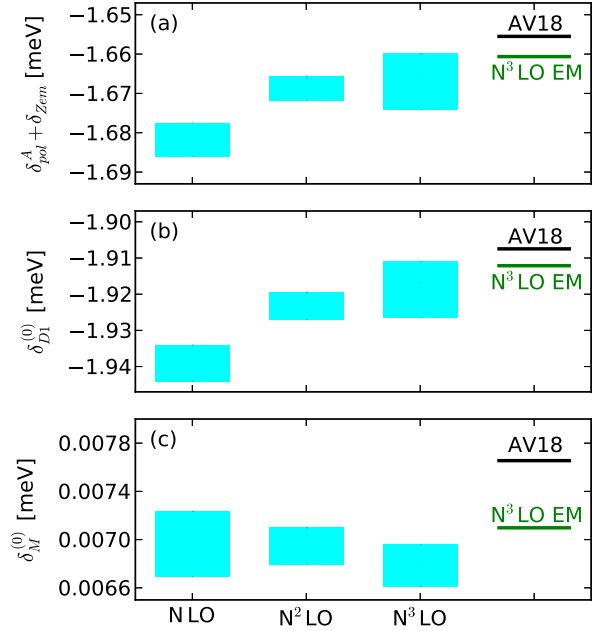}}}
\caption{Total polarization (a), leading order dipole contribution (b) and magnetic dipole contribution (c) at different
chiral orders for the various choices of the cutoffs $\{\Lambda, \tilde{\Lambda}\}$ in comparison to the results with the  AV18 and N$^3$LO-EM potentials.
} 
\label{fig_2_pol} 
\end{figure}

The evolution with EFT orders of 
the total nuclear elastic/inelastic contribution $\delta_{\rm pol}^A+\delta_{\rm Zem}$,
the dipole component $\delta^{(0)}_{D1}$, and the magnetization term
$\delta^{(0)}_{M}$ are presented in Fig.~\ref{fig_2_pol},
using the same values of the cutoffs $(\Lambda, \tilde{\Lambda})$ as in
Table~\ref{table_gs}.
We find similar convergence pattern and cutoff dependence for
$\delta_{\rm pol}^A+\delta_{\rm Zem}$ and $\delta^{(0)}_{D1}$. This is a reflection of the
dipole dominance in $\delta^{A}_{\rm pol}$.
We also note that the cutoff bands increase with the chiral order, similarly to what we observe for 
$\bra r_{str}^2 \ket_d^{1/2}$
and $Q_d$. In fact,  
we found strong correlation between 
$\bra r_{str}^2 \ket_d^{1/2}$
and $\delta^{(0)}_{D1}$ (see also~\cite{Goerke12}).
For the magnetic correction $\delta^{(0)}_{M}$, we observe that the cutoff dependence is large starting from the NLO and it is of comparable size as at the N$^3$LO. This is again due to the effect of the missing two-body currents, which appear already at NLO in the case of magnetic transition~\cite{review}. However, given the size of this term, such further refinements are negligible here.

Comparing the spread of each term in Table~\ref{table_pol} due to the potentials, 
we find that the variation of $\delta^A_{\rm pol}$ with potential models mostly comes from the variation in the dipole term $\delta^{(0)}_{D1}$.
In fact the contribution of all other terms to the spread in $\delta^A_{\rm pol}$ is about $1 \mu\text{eV}$.
Averaging over all potentials we can estimate $\delta^A_{\rm pol}$ as
\begin{equation}
\delta^A_{\rm pol} = -1.239\pm 0.005(1\sigma)\;\;\text{meV}~, 
\end{equation}
and $\delta_{\rm Zem}$ as
\begin{equation}
\delta_{\rm Zem}=-0.424\pm 0.003(1\sigma)\;\;\text{meV}~, 
\end{equation}
where the standard deviation represents an estimate of 
the uncertainty in the high precision nuclear Hamiltonians at a $1\sigma$ level
and can be used to estimate confidence intervals 
for the polarization and the Zemach moment.

From N$^2$LO to N$^3$LO in the $\chi$EFT expansion, the value of $\delta_{\rm Zem}+\delta^A_{\rm pol}$ changes by 0.3\%. This can be considered the systematic error due to the  $\chi$EFT truncation and needs to be included in the total error budget.

The atomic physics error from further corrections
of order $(Z\alpha)^6$ was estimated by Pachucki to be $1\%$~\cite{Pachucki11}. 

Combining all the uncertainties in a quadrature sum, the overall accuracy
of $\delta_{\rm Zem}+\delta^A_{\rm pol}$
is expected to be about 1.16\%.

The proton two-photon exchange contribution to the  $\mu$D Lamb shift is estimated to be $-0.043(3)$ meV~\cite{Pachucki11}. 
However, Ref.~\cite{Carlson14} suggests that only the inelastic part of such contribution is related to $\delta_{\rm pol}^N$. Including  also  the neutron effect, the overall $\delta_{\rm pol}^N$ is estimated as in Ref.~\cite{Carlson14} to be $-0.027(2)$ meV\footnote{Refs.~\cite{Birse12, Birse-14} suggested a different separation of the inelastic part of the nucleon two-photon exchange contributions to the $\mu$H Lamb shift. It shifts correspondingly $\delta^N_{pol}$ in $\mu$D by 0.009 meV, and thus change the central value of Eq.~\eqref{eq:TPE-num} to $-1.681$ meV. The size of this correction is within the uncertainty given in Eq.~\eqref{eq:TPE-num}.}. 
Therefore, we provide a total nuclear/hadron two-photon exchange 
contributions to $\mu$D Lamb shift as
\begin{equation}
\label{eq:TPE-num}
\delta_{\rm TPE} = \delta^A_{\rm pol}+\delta_{\rm Zem} + \delta_{\rm pol}^N = -1.690\pm 0.020\;\;\text{meV}
\end{equation}

The elastic contribution $\delta_{\rm Zem}$ depends on the input of the proton radius. However, due to cancellation of $\delta_{\rm Zem}$ in the elastic and inelastic parts of $\delta_{\rm TPE}$, the only
proton-radius
dependence in $\delta_{\rm TPE}$ comes from the $\delta^{(2)}_{NS}$ term in $\delta^A_{\rm pol}$. Using the CODATA value $0.8775$ fm~\cite{CODATA}, instead of the proton radius $0.8409$ fm from the $\mu$H experiment~\cite{Antognini13}, will change $\delta_{\rm Zem}$ by $-0.007$ meV, but $\delta_{\rm TPE}$ by only $-0.0016$ meV. This is only 0.1\% of $\delta_{\rm TPE}$, thus negligible in the quadrature sum of the error.

\section{Conclusions}

A solid understanding of the theoretical indetermination is crucial
to constrain any explanation of the proton radius puzzle based on physics beyond standard model. In this work we have made an attempt to constrain the nuclear structure corrections
to the $\mu$D Lamb shift utilizing state-of-the-art 
nuclear potentials derived from chiral EFT. Considering the contributions to the nuclear polarization up to order $O(\eta^3)$, where $\eta=\sqrt{m_{\mu}/m_d}$, we obtain an
estimate for $\delta_{\rm TPE}=-1.690$ meV. Combining the spread in the results due to the nuclear potentials ($0.5\%$) and the convergence with $\chi$EFT orders ($0.3\%$), we evaluate 
the nuclear physics error to be about $0.6\%$. This should be added in quadrature to the  $\sim 1\%$ error coming from atomic physics.

Our predicted $\delta_{\rm TPE}$ differs from that obtained by Pachucki \cite{Pachucki11} by only 0.6\%. However, this excellent agreement can be regarded as partly accidental, since differences of several individual contributions are larger than 0.6\%.

Our improved estimates will benefit the determination of 
$\bra {r_{ch}}^2\ket^{1/2}_d$ 
from the ongoing $\mu$D Lamb shift measurement.

\section*{Acknowledgments}

  We thank Hartmuth Arenh\"{o}vel for providing the original data in Ref.~\cite{Arenhoevel}, and Evgeny Epelbaum for offering the $\chi$EFT potential routine. We also thank Franz Kottmann, Judith A. McGovern, and Michael C. Birse for useful discussions.
  This work was supported in part by the Natural Sciences
  and Engineering Research Council (NSERC), the National Research
  Council of Canada, the Israel Science Foundation (Grant number
  954/09), and the Pazy Foundation.


\bibliography{mudeutbib.bib}


\end{document}